\documentclass[twocolumn,prl,aps,showpacs,superscriptaddress,amsmath,amssymb]{revtex4}

\usepackage{graphicx}
\usepackage{amsmath,amsfonts,amssymb}
\usepackage{color}

\newcommand{\dps}{\displaystyle}

\begin{document}

\title{Molecular Simulations of Shock to Detonation Transition in Nitromethane}

\author{Jean-Bernard Maillet}
\author{Emeric Bourasseau}
\affiliation{CEA, DAM, DIF, F-91297 Arpajon, France}

\author{Germain Vallverdu}
\affiliation{\'Equipe de Chimie Physique, IPREM UMR5254, Universit\'e de Pau et des Pays de l'Adour, 2 avenue du Pr\'esident Pierre Angot, 64053 Pau cedex 9, France}

\author{Nicolas Desbiens}
\affiliation{CEA, DAM, DIF, F-91297 Arpajon, France}

\author{Gabriel Stoltz}
\affiliation{Universit\'e Paris Est, CERMICS, Projet MICMAC ENPC - INRIA, 6 \& 8 Av. Pascal, 77455 Marne-la-Vall\'ee Cedex 2, France}

\date{\today}

\begin{abstract}
An extension of the model described in a previous work of Maillet,
Soulard and Stoltz~\cite{maillet07} based on a Dissipative Particule
Dynamics is presented and applied to liquid nitromethane. 
Large scale non-equilibrium simulations of reacting nitromethane under sustained
shock conditions allow a better understanding of the
shock-to-detonation transition in homogeneous explosives. Moreover,
the propagation of the reactive wave appears discontinuous
since ignition points in the shocked material can be activated by the
compressive waves emitted from the onset of chemical reactions.
\end{abstract}

\medskip
\maketitle

%-------------------- ARTICLE ------------------

Since two decades, molecular dynamics (MD) has become a widely
used numerical tool to model and understand shock waves processes. It has been
successfully used to study the appearance of plasticity and phase
transition under shock conditions~\cite{holian98, kadau02}, as well as the structure of the shock
front~\cite{zhakhovskii99}. Similar successes have not been obtained
in the study of high energetic materials, in particular for the
formation of a reactive (detonating) wave. This is mainly due to the
very large time and length scales required to observe such
phenomena. Simulations of reactive waves have been limited so far to
model materials, mainly using the REBO potential or some of its
variants~\cite{brenner93, herring10}. For real energetic
materials, only the first steps of the decomposition mechanism of RDX under
shock conditions were studied using reactive classical
MD~\cite{strachan03}. The same model allowed the simulation of
the onset of the reactive wave~\cite{nomura07}. 
Finally, modified equations of motion (\textit{i.e.} constrained dynamics)
allowed to model shock properties of reacting systems~\cite{reed08}, but the
behavior of the wave itself is not accessible.
In conclusion, full atomistic
simulations of detonations in realistic explosives have never been
achieved so far since the involved computational burden outperforms all current
computer resources by orders of magnitude. 

Having in mind that detonation ultimately is
a multiscale process coupling hydrodynamical features and local chemical reactions, 
coarse graining
strategies appear as a relevant alternative to a full direct numerical simulation of the entire
detonation process. The coarse-graining approach was only recently considered
for shock waves~\cite{holian03,strachan05,stoltz06}. 
It is based on the replacement of complex molecules by
mesoparticules with internal degrees of freedom.

In this letter, we present equations of motions for mesoparticles
including energy exchange between intramolecular and intermolecular
degrees of freedom in order to ensure energy conservation, DPDE
(standing for 'Dissipative Particle Dynamics at constant
Energy'). Moreover, additional variables are added to each
particle to account for its possible chemical evolution. The dynamics
of these variables is governed by classical kinetic equations. The final model, RDPDE
(standing for 'Reactive DPDE'), is used to study the shock to
detonation transition mechanism in
nitromethane.

\paragraph{Description of the model.}
We first briefly recall the DPDE model used to simulate inert 
shock waves in~\cite{stoltz06}. For a system of $N$~particles $q_i$ with momenta
$p_i = m_i v_i$, the equations of motion read

\begin{align}
\dps dq_i = & \dps \frac{p_i}{m_i} \, dt, \nonumber\\
\dps dp_i = & \dps -\nabla_{q_i} V(q) \, dt - \gamma_{ij} \chi^2(r_{ij}) v_{ij} \, dt + \sigma \chi(r_{ij}) \, dW_{ij},\nonumber\\
\dps d\varepsilon_i = & \dps \frac12 \sum_{j, \, j \not = i} \chi^2(r_{ij}) \left (  \gamma_{ij} v_{ij}^2 - \frac{3\sigma^2}{2} \left(\frac{1}{m_i}+\frac{1}{m_j}\right) \right ) \, dt 
\nonumber\\
 & - \sigma \, \chi(r_{ij}) v_{ij} \cdot dW_{ij},
\label{eq:EOM}
\end{align}
where $V(q)$ is the interaction potential between mesoparticules (a sum of paiwise interactions
here), $r_{ij} = |q_i-q_j|$ and $v_{ij} = \frac{p_i}{m_i}-\frac{p_j}{m_j}$ are respectively 
the distance and the relative velocity between particles~$i$ and~$j$, 
$\chi(r)$ is a weighting function with cut-off radius~$r_{\rm cut}$, and the processes $W_{ij}$ are 
independent $d$-dimensional Brownian motions with $W_{ij} = -W_{ji}$. 
The following fluctuation-dissipation relating the magnitude of the stochastic forces $\sigma$ to that of the friction forces $\gamma_{ij}>0$ ensures that the canonical distribution is invariant: $ \gamma_{ij}=\frac{\sigma^2 \beta_{ij}}{2},\quad\beta_{ij} = \frac{1}{2k_B}\left(\frac{1}{T_i}+\frac{1}{T_j}\right)$. In the latter equation, the temperatures~$T_i$ are internal temperatures obtained from an internal equation of state (EOS) on the internal energy: $\varepsilon_i = \int_0^{T_i} C_{v_i}(T) \, dT$. The explicit temperature dependence of the function $C_{v_i}$ allows to accurately reproduce the thermodynamic behavior of the internal degrees of freedom (DoFs). As discussed in~\cite{lynch09}, this function is known to modify the response of materials to shock loading. For reactive material, or in the more general case when some processes are activated by the temperature of internal DoFs, this function is expected to play an important role as well.

For reactive materials, additional variables per particle $\lambda_i$ are introduced to model the progress of the chemical reactions at hand~\cite{maillet07}. Their time evolution equations depend on the order of the chemical reactions and the reversible natures of the latter. We have chosen to model the chemical decomposition of nitromethane with a succession of two first order chemical reactions: The first one is reversible and endothermic, while the second one is irreversible and exothermic. One mesoparticle could represent successively all states beteween NiME and products. The model reaction of the decomposition of nitromethane is:
\[
\mathrm{NiMe} \rightleftarrows \mathrm{NiMe}^* \rightarrow \mathrm{Products}
\]
where NiMe$^*$ represents a metastable material corresponding to a local minimum on the potential energy surface. The progress variables $\lambda_1$ and $\lambda_2$ represent the evolutions of the two chemical reactions, and can be identified with the fractions of NiMe$^*$ 
and Products respectively. 
Their time evolution reads:
\begin{align}
\frac{d\lambda_1}{dt} =& k_1\big(1-\lambda_1-\lambda_2\big)-k_{-1}\lambda_1-k_2\lambda_1,\\
\frac{d\lambda_2}{dt} =& k_2 \, \lambda_1,
\end{align}
where $k_1$ and $k_{-1}$ are the forward and backward reaction rates associated to the first chemical reaction and $k_2$ is the reaction rate associated to the second chemical reaction. The expressions of these rates are given by standard Arrh\'enius expressions : $k = A \exp(-E_a /R\overline{T})$ where $E_a$ is the activation energy and $\overline{T}$ is obtained from a local spacial average of internal temperatures.

The model is constructed so that the total energy of the system (the sum of the 
potential, kinetic, internal and chemical energies) is constant. In particular, 
the energy variations due to chemical reactions are compensated by appropriate variations
of the kinetic and internal energies, as described in~\cite{maillet07}.

\begin{figure}[h!]
\centering
\includegraphics[scale=0.32]{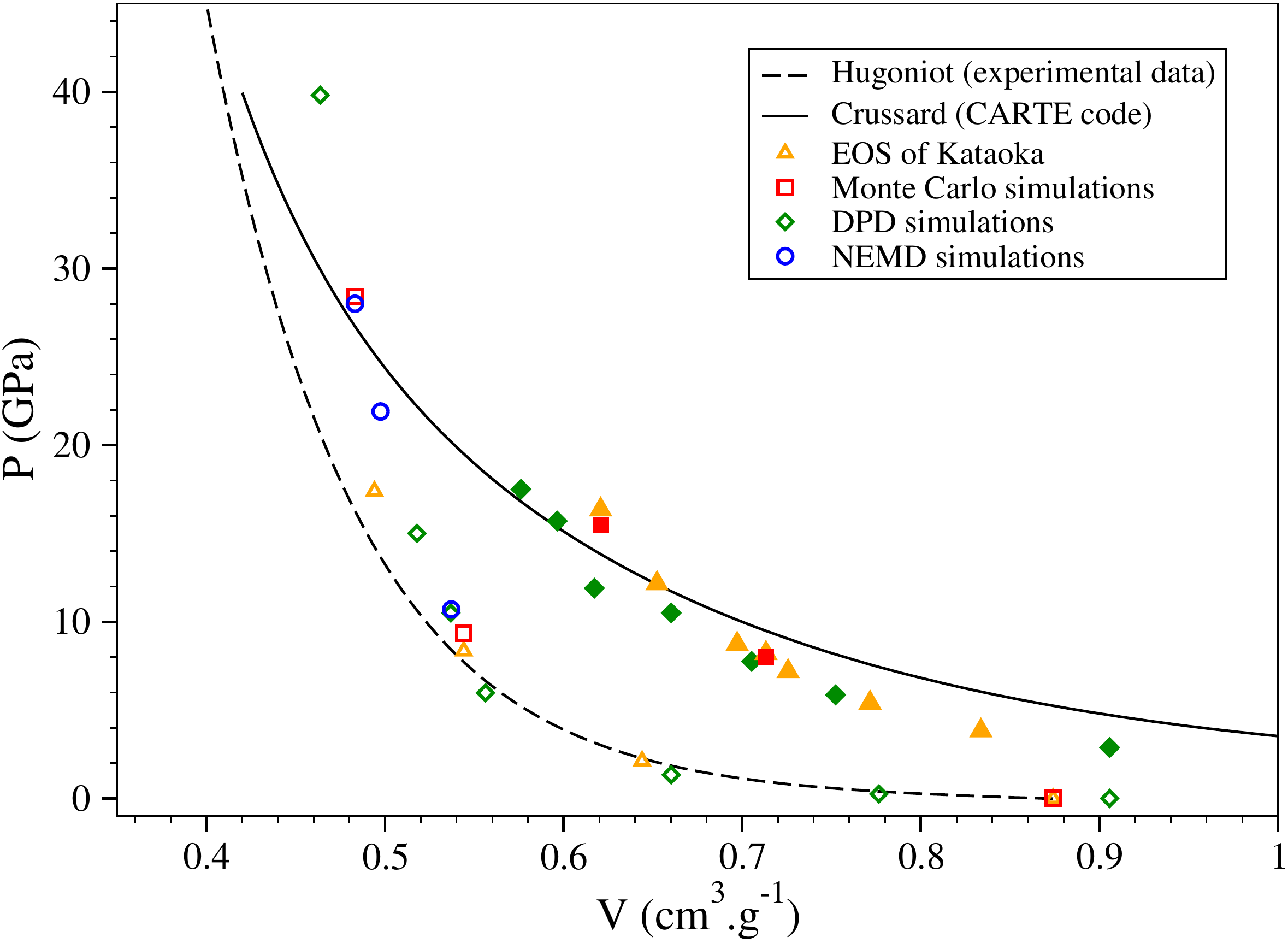}
\caption{Hugoniot and Crussard curves of nitromethane. Black line: experimental Hugoniot curve. Dotted line: Crussard curve (CARTE code). Open symbols correspond to inert nitromethane while full symbols correspond to detonation products. Triangles: calculations made with the EOS of Kataoka. Squares: Monte Carlo simulations. Diamonds: MD-DPD equilibrium simulations. Circles: non equilibrium shock simulations.}
\label{fig:hugoniot}
\end{figure}

\paragraph{Application to nitromethane.}
The interactions between mesoparticules are described by a classical
exp-6 potential, which is known to ensure a good compressibility of
the system at high pressure. By means of the analytical expression of
the equation of state of exp-6 fluids proposed by
Kataoka~\cite{kataoka92}, the parameters of the potential for 
inert nitromethane and its detonation products
are obtained respectively by optimization on experimental
Hugoniot~\cite{Marsh,klebert98,desbiens09} and Crussard
curves~\cite{dubois10}. Consistency of these potentials
with experimental data is shown on Figure \ref{fig:hugoniot}. From now
on, computations are made with these potentials. When a DPD particule
is reacting ($0 < \lambda_i < 1$), the parameters of the interaction
potential are given by a linear interpolation of the parameters of the
potential for the inert and the fully reacted system.

For inert nitromethane, as the DPD particle represents only a single
nitromethane molecule, there is a clear splitting between internal and
external DoFs: the internal $C_v$ of the DPD particle represents
simply the sum of the intramolecular contributions to the
energy. Hence the $C_v$ function is taken directly from thermodynamic
tables. For the fully reacted material, a DPD particle represents the
decomposition products of one nitromethane molecule, \textit{i.e.} a
group of several small molecules. The corresponding internal $C_v$ is
then a sum of two contributions: the internal DoFs of the small
molecules plus a fraction of their external DoFs. There is no clear
splitting between intra and intermolecular DoFs in the coarse graining
process. Therefore, the $C_v$ function of the DPD particle cannot be
extracted from a standard thermodynamic table, and has to be computed
numerically as the difference between the total $C_v$ of the real
system and the $C_v$ due to the intermolecular DPD potential only. In
the following internal DoFs will stands for all DoFs inside a
mesoparticle and external DoFs will stands for displacement of
mesoparticles.

The rate of heat exchanges between internal and external DoFs is
controlled by the parameter~$\sigma$. Dawes \emph{et al}~\cite{dawes09} 
have presented a study of energy relaxation in
solid nitromethane behind a shock wave. They computed the rate of
energy transfer between translational and rotational kinetic energies
and internal vibrational modes. Of course, some of the vibrational
modes that exhibit an efficient coupling to the phonons are therefore heated
rapidly after the shock, while some other modes heat up more
slowly. In the DPD model we use, there is only one mode representing the
average of all internal vibrations. We chose a value of
$\sigma$ leading to an equilibration time of a few picoseconds
between intra and intermolecular DoFs.

\paragraph{Numerical results.}
All simulations have been performed with our in-house parallel MD code.
We used a time step $\Delta t = 1$~fs in our simulations, a value 
at least one order of magnitude larger than for MD simulations with 
standard reactive potentials as such ReaxFF.

The thermodynamic properties of our model have been studied by 
equilibrium and nonequilibrium MD simulations. 
The computed density of the material at $P = 1$~bar and $T = 300$~K is 
1.1041g.cm$^{-3}$. 
Simulations of inert nitromethane using the hugoniostat constraint
method and NEMD (shock simulations) have been performed in order to compute
the inert Hugoniot curve. Similar simulations are performed with
reacted nitromethane to compute the Hugoniot curve of
detonation products, \textit{i.e.} the Crussard curve. Results are displayed in
Figure~\ref{fig:hugoniot}.
A semi-quantitative agreement is found on thermodynamical properties of both the 
inert and reacted explosive.

Nonequilibrium shock wave simulations have been performed with the RDPDE model. 
Sustained shock simulations have been loaded on sample of size 20$\times$20$\times$10000 unit cells (about 4.5$\mu$m long), with an infinitely massive piston moving at constant velocities equal to 1500, 2000, 2500, 2700 and 3000~m.s$^{-1}$. The total simulation time for each simulation is around 700~ps. In the simulations with piston velocities equal to 1500 and 2000~m.$^{-1}$, no sign of reactive wave formation is observed. The inert shock wave propagates through the whole sample, activating the explosive with the first reversible reaction, but the detonation products are not produced. This is due to the fact that the shock did not release enough energy for the system to overcome the energy barrier of the second (irreversible) reaction. For piston velocities equal to or higher than 2500~m.s$^{-1}$, we observe the decomposition of the explosive and the formation of a reactive wave. The velocity of the corresponding wave now exceeds the one of the inert shock wave due to the global energy release of exothermic reactions. 

On the experimental side, multiple-magnetic gauge measurements of neat nitromethane~\cite{sheffield06} allowed the direct observation of the shock-to-detonation mechanism: A reactive wave first builds up in the shocked (inert) material (characterized by an increase of the particle velocity as the wave propagates) and forms a superdetonation wave travelling at constant velocity. This superdetonation wave overtakes the inert shock wave and progressively decays to an overdriven, and eventually to a steady, detonation. This has recently been confirmed by PDV (Photonic Doppler Velocimetry) measurements~\cite{mercier10}.

\begin{figure}[h!]
\centering
\includegraphics[scale=0.32]{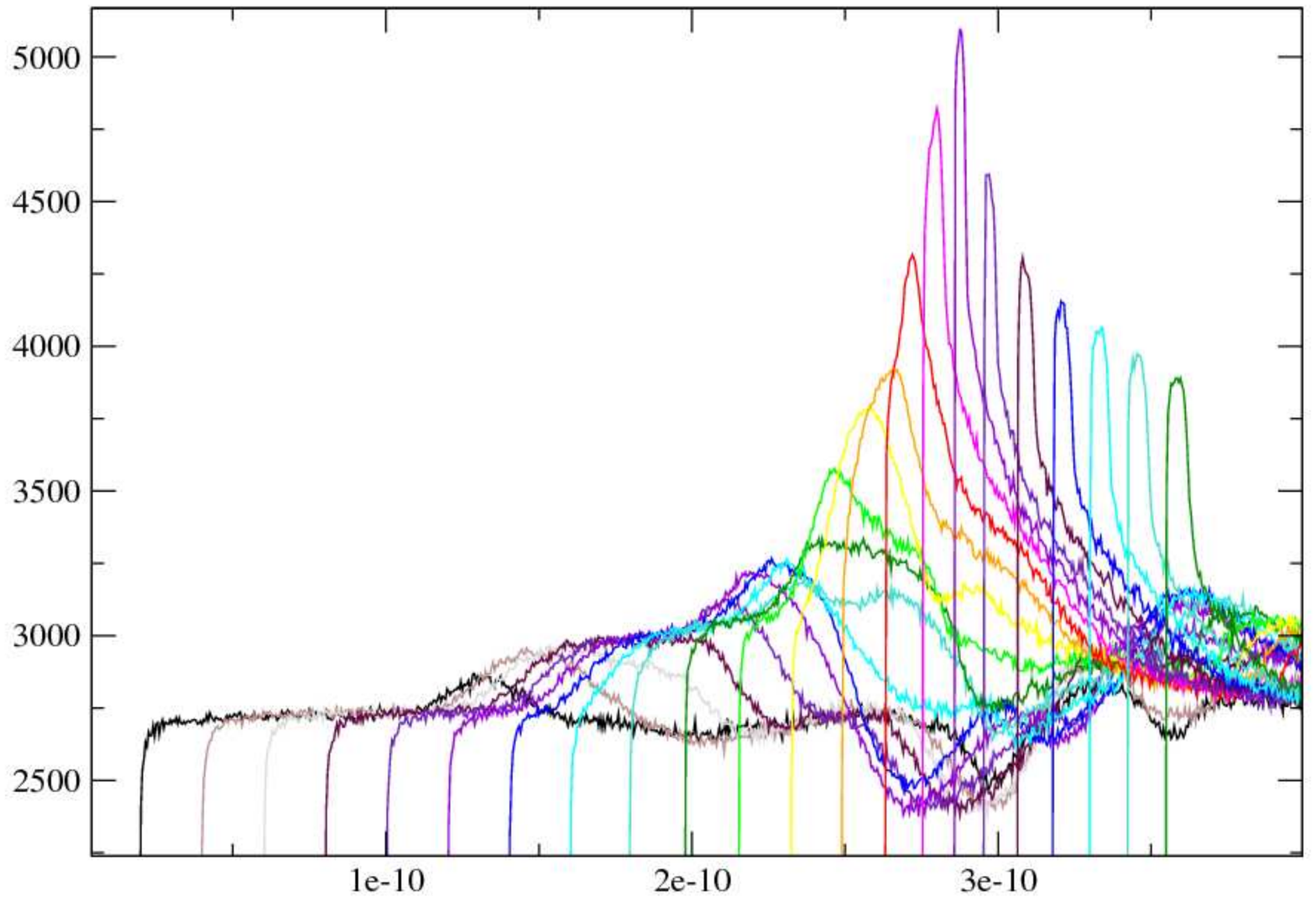}
\caption{Time evolutions of particle velocities taken at different locations in the material (every 50 cells), mimicking the behaviour of multiple gauges set up.}
\label{fig:profile_up_t}
\end{figure}

\begin{figure}[h!]
\centering
\includegraphics[scale=0.5]{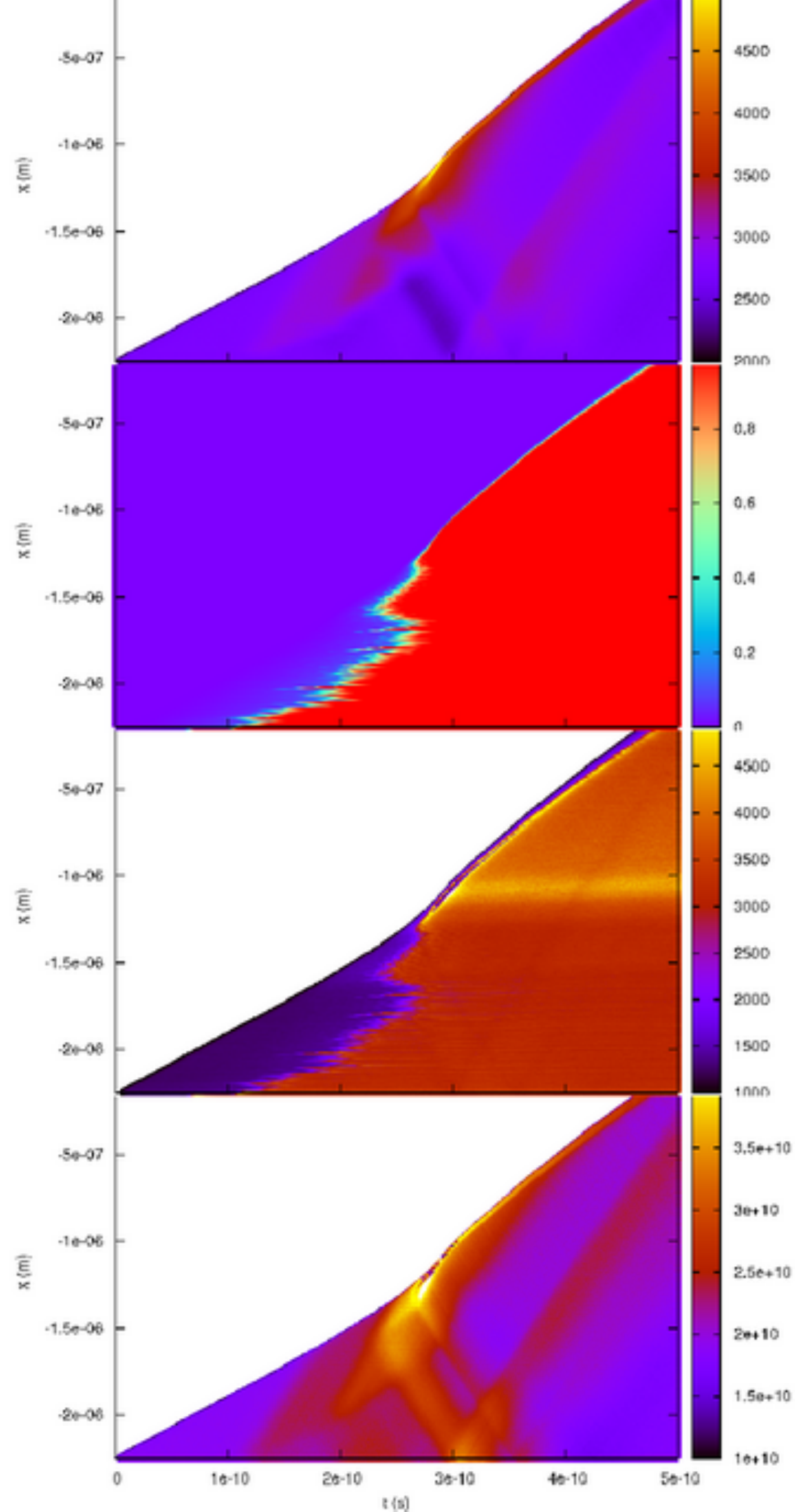}
\caption{Coloured maps of thermodynamic variable (particle velocity, progress variable, internal temperature and pressure) in a time-space diagram.}
\label{fig:map_totale}
\end{figure}

 For the simulation using a piston velocity of 2700~m.s$^{-1}$, the onset of the reactive wave is evidenced in Figure~\ref{fig:profile_up_t} by the signal of (artificial) multiple gauges.
This wave progressively catches up with the shock wave. These signals exhibit striking similarities with experimental measurements, although occuring on a shorter timescale. The complete scenario of the shock-to-detonation transition is displayed in Figure~\ref{fig:map_totale}, which shows the evolutions of the average velocity, progress variable, internal temperature and the $x$-component of the pressure tensor, in a $(x,t)$~diagram.

After a shock wave is loaded in the sample it leaves the nitromethane
in an shocked state. After around 100~ps, chemical reactions
start to occur at the interface with the piston since this is the
region of space where the system has spent the longest time in the
shocked state. As the chemical reactions progress, a reactive wave is
formed, which travels in the shocked nitromethane at a higher velocity
than the shock wave. Its front is given by the analysis of the
diagram of the progress variable. From the onset of chemical reactions
happening around 100~ps, a compressive wave is created, travelling
much faster than the reactive wave in the shocked material, up to the
inert shock front. This compressive wave brings the inert shocked
system at a slightly higher density. When it reaches the shock front,
it accelerates and increases the pressure (and the material velocity)
in the shocked inert material. At some point, an ignition zone induced
by the previously described compressive wave appears in the region
between the inert shock wave and the reactive wave. This ignition
point acts as a hot spot, and chemical reactions develop from this
point forward and backward in the material, forming two reacting
waves. Once again, these reacting waves are preceded by compressive
waves. A careful analysis of the diagram reveals that a second hot
spot is activated, closer to the front shock. As the reactive wave
develops from this point, it catches up with the shock wave, forming a
single reactive wave, the overdriven detonation. This overdriven
detonation propagates at high speed and is characterized by high
particle velocities, as well as high values of thermodynamic variables
(temperature, pressure,...). Once formed, the overdriven detonation
progressively decays to a stationary detonation, as evidenced in the
pressure and temperature diagrams.

In conclusion, we presented for the first time a molecular simulation
of a shock-to-detonation transition in a realistic energetic material,
using a reduced model based on a DPD coarse graining approach. Our
numerical results confirmed the shock to detonation mechanism observed
experimentally for nitromethane. Moreover, we have shown that
compressive waves are emitted from the onset of chemical reactions,
therefore increasing the probability of ignition forward in the
material. Hence the reactive wave exhibits a discontinuous progression  from sites to sites up to the shock front.

\acknowledgments{
Laurent Soulard is gratefully acknowledged for fruitful discussions. G.S. is supported in part by the Agence Nationale de la Recherche, 
  under the grant ANR-10-BLAN 0108 (MEGAS).
All Monte Carlo simulations have been performed with the GIBBS code from IFP, CNRS and the Universit\'e Paris-Sud~\cite{ungerer}.
}


\begin{thebibliography}{10}

\bibitem{maillet07}
J.-B. {\textsc{Maillet}}, L.~{\textsc{Soulard}}, and G.~{\textsc{Stoltz}}.
\newblock {\em Euro. Phys. Lett.}, 78:68001, 2007.

\bibitem{holian98}
B.L. {\textsc{Holian}} and P.S. {\textsc{Lomdahl}}.
\newblock {\em Science}, 280:2085, 1998.

\bibitem{kadau02}
K.~{\textsc{Kadau}}, T.C. {\textsc{Germann}}, P.S. {\textsc{Lomdahl}}, and B.L.
  {\textsc{Holian}}.
\newblock {\em Science}, 296:1681, 2002.

\bibitem{zhakhovskii99}
V.V. {\textsc{Zhakhovskii}}, S.V. {\textsc{Zybin}}, K.~{\textsc{Nishihara}},
  and S.I. {\textsc{Anisimov}}.
\newblock {\em Phys. Rev. Lett.}, 83:1175, 1999.

\bibitem{brenner93}
D.W. {\textsc{Brenner}}, D.H. {\textsc{Robertson}}, M.L. {\textsc{Elert}}, and
  C.T. {\textsc{White}}.
\newblock {\em Phys. Rev. Lett.}, 70:2174, 1993.

\bibitem{herring10}
D.~{\textsc{Herring}}, T.C. {\textsc{Germann}}, and
  N.~{\textsc{Gronbech-Jensen}}.
\newblock {\em Phys. Rev. E.}, 82:214108, 2010.

\bibitem{strachan03}
A.~{\textsc{Strachan}}, A.C.T. {\textsc{van Duin}}, D.~{\textsc{Chakraborty}},
  S.~{\textsc{Dasgupta}}, and W.A. {\textsc{Goddard}}.
\newblock {\em Phys. Rev. lett.}, 91(9):098301, 2003.

\bibitem{nomura07}
K.I. {\textsc{Nomura}}, R.K. {\textsc{Kalia}}, A.~{\textsc{Nakano}},
  P.~{\textsc{Vashishta}}, A.C.T. {\textsc{van Duin}}, and W.A.
  {\textsc{Goddard}}.
\newblock {\em Phys. Rev. Lett.}, 88:148303, 2007.

\bibitem{reed08}
E.J. {\textsc{Reed}}, M.R. {\textsc{Manaa}}, L.E. {\textsc{Fried}}, K.R.
  {\textsc{Glaesemann}}, and J.D. {\textsc{Joannopoulos}}.
\newblock {\em Nature Physics}, 4:72, 2008.

\bibitem{holian03}
B.L. {\textsc{Holian}}.
\newblock {\em Europhys. Lett.}, 64:330, 2003.

\bibitem{strachan05}
A.~{\textsc{Strachan}} and B.L. {\textsc{Holian}}.
\newblock {\em Phys. Rev. lett.}, 94:014301, 2005.

\bibitem{stoltz06}
G.~{\textsc{Stoltz}}.
\newblock {\em Europhys. Lett.}, 76:849, 2006.

\bibitem{lynch09}
K.~{\textsc{Lynch}}, A.~{\textsc{Thompson}}, and A.~{\textsc{Strachan}}.
\newblock {\em Modelling Simul. Mater. Sci. Eng.}, 17:015007, 2009.

\bibitem{kataoka92}
Y.~{\textsc{Kataoka}}.
\newblock {\em Bull. Chem. Soc. Jpn.}, 65:2093, 1992.

\bibitem{Marsh}
Marsh.
\newblock {\em LASL shock hugoniot data}.
\newblock 1980.

\bibitem{klebert98}
P.~{\textsc{Klebert}}.
\newblock {\em Etude exp\'erimentale et th\'eorique de la transition choc
  d\'etonation dans les explosifs homog\`enes}.
\newblock PhD thesis, Paris VI - Pierre et Marie Curie University, 1998.

\bibitem{desbiens09}
N.~{\textsc{Desbiens}}, E.~{\textsc{Bourasseau}}, J.-B. {\textsc{Maillet}}, and
  L.~{\textsc{Soulard}}.
\newblock {\em J. of Hazardous Materials}, 166:1120, 2009.

\bibitem{dubois10}
V.~{\textsc{Dubois}}, N.~{\textsc{Desbiens}}, and E.~{\textsc{Auroux}}.
\newblock {\em Chem. Phys. Lett.}, 494:306, 2010.

\bibitem{dawes09}
R.~{\textsc{Dawes}}, A.~{\textsc{Siavosh-Haghighi}}, T.D. {\textsc{Sewell}},
  and D.L. {\textsc{Thompson}}.
\newblock {\em J. Chem. Phys.}, 131:224513, 2009.

\bibitem{sheffield06}
S.A. {\textsc{Sheffield}}, D.M. {\textsc{Dattelbaum}}, R.~{\textsc{Engelke}},
  R.R. {\textsc{Alcon}}, B.~{\textsc{Crouzet}}, D.L. {\textsc{Robbins}}, D.B.
  {\textsc{Stahl}}, and R.L. {\textsc{Gustavsen}}.
\newblock In proceedings of the 13th Symposium of Detonation, 2006.

\bibitem{mercier10}
P.~{\textsc{Mercier}}, J.~{\textsc{B\'enier}}, P.A. {\textsc{Frugier}},
  M.~{\textsc{Debruyne}}, and B.~{\textsc{Crouzet}}.
\newblock New models and Hydrocodes for shock wave processes in Condensed
  Matter, Paris, EPJ web of conference \textbf{10}, 00016 (2010), 2010.

\bibitem{ungerer}
P.~{\textsc{Ungerer}}, A.~{\textsc{Boutin}}, and B.~{\textsc{Tavitian}}.
\newblock {\em Applications of Molecular Simulation in the Oil and Gas
  Industry}.
\newblock IFP Publications, 2005.

\end{thebibliography}
\end{document}